\documentclass{optica-article}

\journal{opticajournal} 

\articletype{Research Article}

\usepackage{lineno}

\usepackage{amsmath,bm}
\usepackage{graphicx}
\usepackage{booktabs}
\usepackage{multirow}
\usepackage{tabularx}
\usepackage{siunitx}
\usepackage{float}
\usepackage{hyperref}

\begin{document}

\title{X-ray phase-contrast imaging featuring a tabletop laser-driven X-ray source operating in ambient conditions}

\author{Andrew Coathup\authormark{1,2,3,*}, Aarón Alejo\authormark{3}, Antón Barreiro-Iglesias\authormark{4,5}, Laura González-Llera\authormark{4,5}, Alicia Reija\authormark{3}, and José Benlliure\authormark{6}}

\address{
\authormark{1}Henry Royce Institute, Department of Materials, University of Manchester, Royce Hub Building, Oxford Rd, M13 9PL, Manchester, UK\\
\authormark{2}National Facility for X-ray Computed Tomography (NXCT), University of Manchester, Royce Hub Building, Oxford Rd, M13 9PL, Manchester, UK\\
\authormark{3}Instituto Galego de Física de Altas Enerxías (IGFAE), Universidade de Santiago de Compostela, Santiago de Compostela, Spain\\
\authormark{4}Departamento de Bioloxía Funcional, Facultade de Bioloxía, Universidade de Santiago de Compostela, 15782, Santiago de Compostela, Spain\\
\authormark{5}Aquatic One Health Research Center (ARCUS), Universidade de Santiago de Compostela, 15782, Santiago de Compostela, Spain\\
\authormark{6}Instituto de Física Corpuscular (IFIC), CSIC- Universitat de València, Valencia, Spain
}

\email{\authormark{*}andrew.coathupwhale@manchester.ac.uk} 


\begin{abstract*}
Propagation-based X-ray phase-contrast imaging (PBI) is demonstrated using the recently commissioned tabletop femtosecond laser-driven X-ray source at the Laser Laboratory for Acceleration and Applications hosted at the University of Santiago de Compostela. The imaging conditions for PBI were systematically investigated using phase-induced edge enhancement as a figure of merit. Edge enhancement increased strongly with increasing object-to-detector distance and remained strong at high geometric magnifications, reaching a maximum value of 2.2\% at a source-to-detector distance of 63~cm and a magnification of $35\times$. The compact, high magnification imaging setup was possible due to the source's operation in ambient conditions without the need for a vacuum chamber, allowing for greater flexibility in sample positioning. The chosen imaging setup conditions were then used for high-magnification ($42\times$) imaging of a zebrafish at 20 days post fertilization (dpf), demonstrating the potential of the laser-driven X-ray source for PBI of small biological samples, including those that may be incompatible with vacuum environments.
\end{abstract*}


\section{Introduction}

X--ray phase-contrast imaging (XPCI) is an advanced X-ray imaging technique that uses the changes in the phase of X-rays as they pass through a sample to improve image quality. Compared with conventional attenuation-based X-ray imaging, XPCI can provide improved contrast between weakly absorbing materials, and has found application in fields such as materials science~\cite{eastwood2014application}, cultural heritage~\cite{mocella2015revealing}, and virtual histology~\cite{donato2026spectral}.

There are several approaches to perform XPCI, including interferometric~\cite{momose2002phase}, analyser-based~\cite{morgan2012x}, and modulation-based techniques using speckles~\cite{zanette2014speckle}, gratings~\cite{pfeiffer2013grating}, or masks~\cite{olivo2021edge}. However, these approaches generally require more involved experimental setups with additional optical or mechanical components. Propagation-based X-ray phase-contrast imaging (PBI)~\cite{gureyev2009refracting}, conversely, relies only on the free-space propagation of the X-ray wavefront between the object and detector, producing characteristic edge enhancement fringes at material interfaces~\cite{dierks2020experimental}. PBI therefore offers a comparatively simple experimental setup, however it does require sufficient lateral spatial coherence, and consequently a sufficiently small effective X-ray source, to preserve the phase effects~\cite{endrizzi2018x}.

Due to these requirements, PBI has traditionally been performed using synchrotron radiation, where very high photon flux and strong spatial coherence provide excellent conditions for X-ray phase-contrast imaging~\cite{cloetens1999hard}. However, limited access to synchrotron facilities has motivated the development of laboratory-scale alternatives. X-ray imaging laboratories often employ microfocus X-ray tubes, which provide the small source dimensions required for PBI~\cite{wilkins1996phase}, but the achievable photon flux is limited by thermal loading of the target, often resulting in extensive acquisition times.

There is therefore considerable interest in compact X-ray sources that combine the small source dimensions and high photon flux desirable for laboratory PBI. Liquid-metal-jet sources are one alternative~\cite{trapani2024speckle}, overcoming the thermal-loading limitations of conventional microfocus sources by continuously refreshing the target material. Another promising approach are laser-driven X-ray sources, offering a micrometre-scale source dimension and higher photon fluxes, while fitting within a laboratory-scale system.

Several types of laser-driven X-ray sources have been investigated for PBI. Laser-wakefield accelerators produce bright betatron radiation from small emission regions and have been demonstrated for X-ray phase-contrast imaging of biological samples~\cite{kneip2011x,cole2018high,wenz2015quantitative}, while laser-driven inverse-Compton sources can generate compact, high-brightness X-ray beams~\cite{gunther2020versatile}. Alternatively, intense ultrashort laser pulses interacting with solid targets can generate X-rays from micrometre-scale sources through bremsstrahlung and characteristic K-$\alpha$ emission, allowing for PBI of biological and weakly absorbing samples~\cite{gambari2020exploring}. Laser-plasma sources have also enabled PBI of rapidly evolving systems such as laser-driven shock waves~\cite{barbato2019quantitative}. However, to the best of the authors' knowledge, previous demonstrations of PBI using laser-driven X-ray sources have all relied on vacuum environments for X-ray generation, imposing physical constraints on the experimental setup, particularly for compact, high-magnification imaging. Laser-driven X-ray sources operating in ambient conditions could reduce these constraints and offer an interesting alternative for laboratory-based PBI.

Such a laser-driven X-ray source~\cite{reija2025optimisation} has recently been developed at the Laser Laboratory for Acceleration and Applications (L2A2)~\cite{benlliure2019validation}, hosted at the University of Santiago de Compostela, Spain. The X-ray source is driven by low-energy ($<1$~mJ), 35~fs laser pulses at a repetition rate of 1~kHz, which pass freely through air before interacting with a rotating solid copper target to generate X-rays. The resulting micrometre-scale X-ray source dimensions and high photon flux provide favourable characteristics for PBI, while operation in ambient conditions offers greater experimental flexibility, including short source-to-object distances (SODs) and large geometric magnifications within a compact tabletop imaging geometry.

In this work, we investigate the implementation of PBI using the laser-driven X-ray source at the L2A2. The effects of source-to-detector distance (SDD), object-to-detector distance (ODD), and geometric magnification on phase-induced edge enhancement are systematically evaluated. The influence of surrounding material on the visibility of phase-contrast is also investigated. Finally, the chosen imaging setup conditions are applied to PBI of zebrafish at high geometric magnification.

\section{Materials and Methods}

\subsection{Laser-driven X-ray source}

A tabletop laser-driven X-ray source operating in ambient conditions has been commissioned at the L2A2 hosted at the University of Santiago de Compostela~\cite{reija2025optimisation}. The source is driven by the low-energy beamline of an ALPHA 10/XS Ti:sapphire laser system (Thales), which delivers pulses with a central wavelength of 800~nm at a repetition rate of 1~kHz~\cite{benlliure2019validation}. The laser pulse energy measured at the entrance to the experimental area is 0.8~mJ, with a pulse duration of 35~fs at full width at half maximum (FWHM).

The p-polarized laser pulses propagate through air and are focused by a Mitutoyo Plan Apo NIR $20\times$ microscope objective onto a 1~mm thick rotating copper target oriented at $45^\circ$ with respect to the incident laser beam. The laser focal spot was measured to be 1.5~$\mu$m (FWHM), with an estimated peak intensity of $1.4\times10^{17}$~W~cm$^{-2}$ at the target~\cite{reija2025optimisation}. The target is mounted on a motorized positioning system that maintains the target surface within the short focal region (Rayleigh length) of the laser ($\sim$~8.4~$\mu$m) while continuously providing a fresh interaction surface for successive laser pulses, using a variation of the correction described in Ref.~\cite{penas2024multi}.

The laser--target interaction produces characteristic Cu K-$\alpha$ X-ray emission at approximately 8~keV together with a bremsstrahlung continuum. As discussed in Ref.~\cite{reija2025optimisation}, the Cu K-$\alpha$ photon flux was measured to be $2.3\times10^{9}$~photons~s$^{-1}$ over $2\pi$~sr, while the X-ray source dimensions were measured to be approximately $12\times8~\mu\mathrm{m}^{2}$~\cite{reija2025optimisation}.

The combination of micrometre-scale source dimensions, high X-ray flux, and good operational stability makes the source well suited to PBI. The small source size limits geometric unsharpness and helps preserve the spatial coherence required for propagation-induced phase effects. The high repetition rate and X-ray flux allow sufficient photon statistics to be accumulated within practical laboratory acquisition times. Furthermore, operation in ambient conditions simplifies the experimental setup and helps limit the deposition of copper debris from the laser--target interaction onto the microscope objective. It also allows samples to be positioned close to the X-ray generation region, providing access to the short source-to-object distances and high geometric magnifications investigated in this work.

\subsection{PBI imaging setup}

\begin{figure}[htbp]
    \centering
    \includegraphics[width=\linewidth]{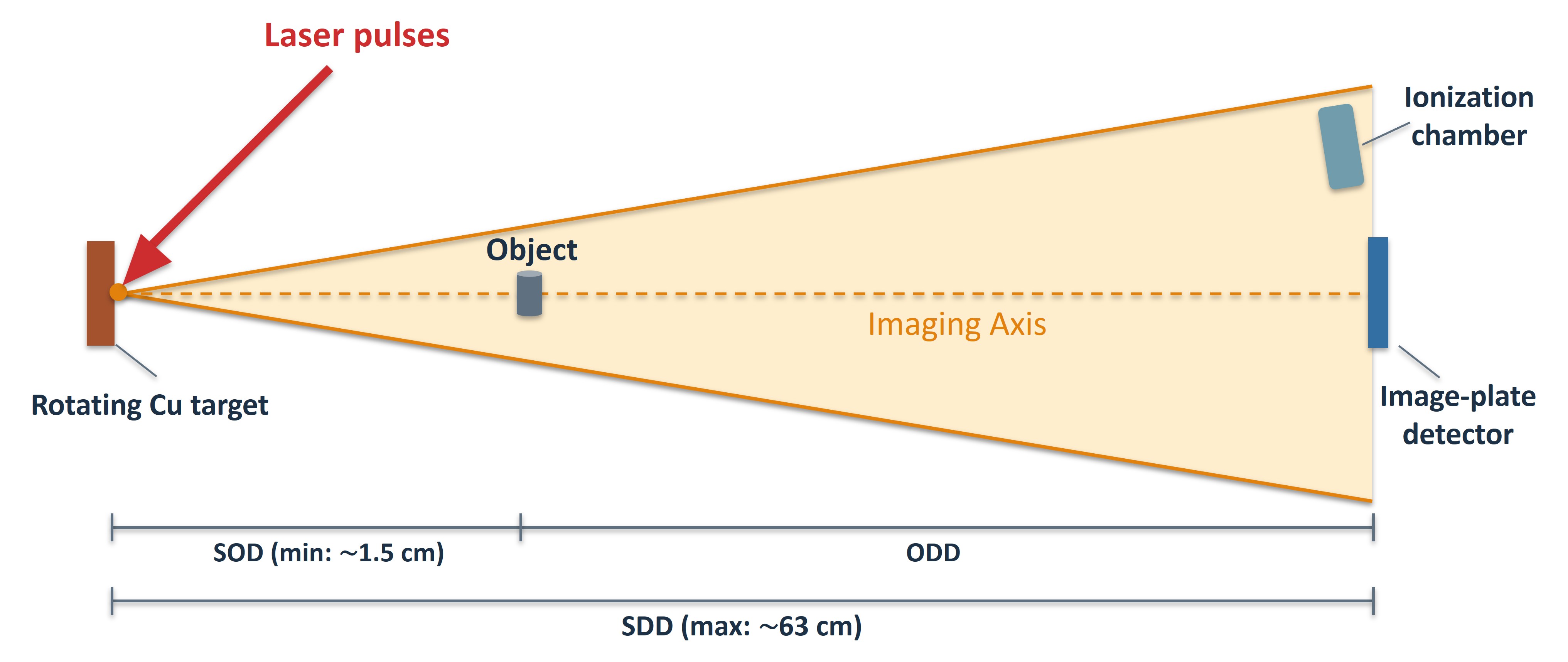}
    \caption{Schematic overview of the experimental PBI setup. The red arrow indicates the incident laser pulses, while the dashed yellow line indicates the X-ray imaging axis. The minimum achievable SOD is approximately 1.5~cm, while the maximum SDD is approximately 63~cm.}
    \label{fig:setup}
\end{figure}

The PBI imaging setup was assembled on a $90~\mathrm{cm}\times60~\mathrm{cm}$ optical table, with the principal components shown in Fig.~\ref{fig:setup}. The laser pulses are directed into the Mitutoyo $20\times$ microscope objective, which focuses the beam onto the rotating copper target to drive the X-ray emission. An ionization chamber (PTW 786) positioned adjacent to the imaging axis was used to monitor the X-ray output during irradiation and ensure consistency between subsequent irradiations.

The imaging axis, indicated by the yellow dashed line in Fig.~\ref{fig:setup}, extends from the X-ray generation site along the length of the optical table, providing a maximum SDD of approximately 63~cm. The object and detector could be freely and independently positioned along this axis, enabling the imaging geometry to be varied systematically. Depending on the dimensions and mounting of the object, a minimum SOD of $\sim$1.5~cm could be achieved, allowing for geometric magnifications exceeding $40\times$. The geometric magnification, $M$, is given by:

\begin{equation}
    M=\frac{\mathrm{SDD}}{\mathrm{SOD}}.
    \label{eq:magnification}
\end{equation}

The X-ray detector used throughout the experiments described in this work was a photostimulable phosphor dental image plate (D\"urr Dental), with a field of view of $6~\mathrm{cm}\times8~\mathrm{cm}$ and a 16-bit grayscale readout~\cite{durrDental}. Following each X-ray exposure, the image plate was digitized using the corresponding image plate scanner with an effective pixel size of 22~$\mu$m.

\subsection{Experimental design and simulations}

A nylon string was selected as a simple and inexpensive phase-contrast phantom to investigate the PBI setup geometry for the laser-driven X-ray source. Nylon produces phase-induced edge enhancement at its boundaries with air and is well suited to the relatively low-energy X-ray spectrum produced by the source. Moreover, its cylindrical geometry provides two well-defined interfaces from which phase-induced edge enhancement can be quantified.

To investigate the dependence of phase-induced edge enhancement on imaging geometry for the laser-driven X-ray source setup, numerical simulations were performed using a simplified version of a previously validated X-ray phase-contrast simulation tool~\cite{coathup2025simulation}. A 500~$\mu$m-diameter nylon cylinder was modelled using a monochromatic photon energy of 8~keV, approximately corresponding to the characteristic Cu K-$\alpha$ emission of the laser-driven X-ray source. The complex refractive index of nylon was retrieved automatically through the simulation tool via XrayDB~\cite{xraydb}, assuming a chemical composition of $\mathrm{C_{12}H_{22}N_2O_2}$ and a density of 1.14~g~cm$^{-3}$. The simulations also assumed an X-ray source size of 8~$\mu$m, a simplified image plate detector response with a pixel size of 22~$\mu$m, and the absence of image noise.

Different metrics can be used to quantify phase-induced edge enhancement~\cite{dierks2020experimental,monnin2004quantitative,sung2017phase}. In the present work, edge enhancement was quantified by considering the relative increase in intensity at the positive edge peak compared to the background intensity:

\begin{equation}
    EE=\frac{I_E-I_{\mathrm{bkg}}}{I_{\mathrm{bkg}}},
    \label{eq:edge_enhancement}
\end{equation}

where $I_E$ is the maximum intensity at the edge peak and $I_{\mathrm{bkg}}$ is the corresponding background intensity. As the simulations did not include noise, the background intensity was determined directly from the simulated intensity profile and was equal to 1. Edge enhancement values are expressed as percentages throughout this work. 

\begin{figure}[htbp]
    \centering
    \includegraphics[width=\linewidth]{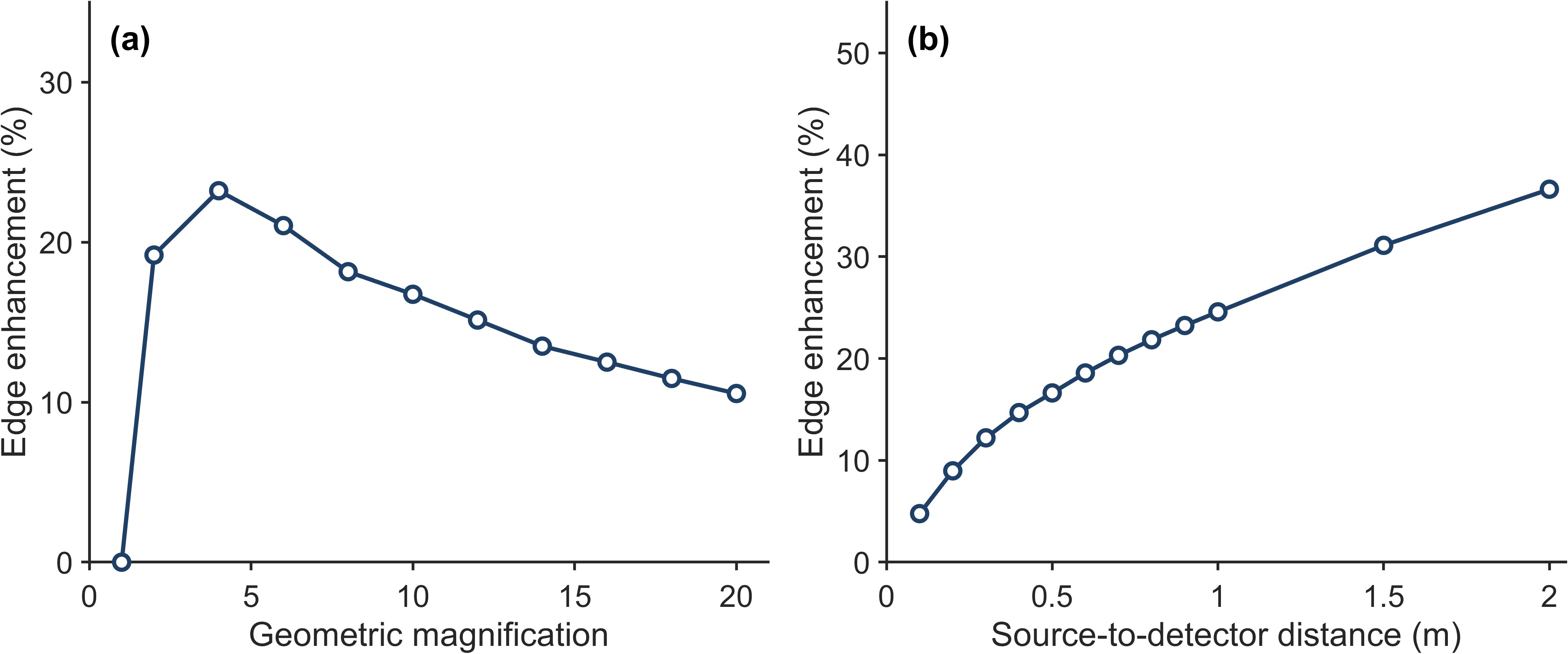}
    \caption{Simulated dependence of phase-induced edge enhancement on (a) geometric magnification at a fixed SDD of 44~cm and (b) SDD at a fixed magnification of $11\times$.}
    \label{fig:simulations}
\end{figure}

The SDD and geometric magnification were varied independently to investigate their relative influences on phase-induced edge enhancement. The SDD was varied between 0.1~m and 2.0~m at a fixed magnification of $11\times$, while the magnification was varied between $1\times$ and $20\times$ at a fixed SDD of 44~cm. The reference values of 44~cm and $11\times$ were chosen as they corresponded to a readily accessible imaging geometry that did not interfere with other components on the optical table. Figure~\ref{fig:simulations} shows the results of these simulations, illustrating the dependence of phase-induced edge enhancement on imaging geometry under the assumed conditions. The calculated edge enhancement increased strongly with SDD over the simulated range, which extended well beyond the maximum SDD accessible on the optical table. In contrast, increasing the geometric magnification produced a rapid increase in the calculated edge enhancement, reaching a maximum at an intermediate magnification of approximately $4\times$, followed by a gradual decrease at higher magnifications. These results show that large SDDs and a minimum geometric magnification are important for PBI. However, in practice, the choice of magnification will also depend on the sample dimensions and the required field of view, while the maximum achievable SDD will also depend on the available X-ray flux and acceptable exposure times.

\subsection{Experimental image analysis and phase-contrast metrics}

The influence of imaging geometry on phase-induced edge enhancement was investigated experimentally using SDDs of 30~cm, 44~cm, and 63~cm, with geometric magnifications ranging from approximately 1 to 35$\times$, varied by adjusting the SOD. Additionally, the dependence on ODD was investigated at a fixed SOD of 10~cm, with ODDs of 20~cm, 30~cm, 41~cm, and 52~cm. For the magnification scans, the accumulated ionization chamber charges were maintained at approximately 0.2~nC, 0.5~nC, and 1.5~nC for SDDs of 30~cm, 44~cm, and 63~cm, respectively, corresponding to irradiation times from 28 to 58~s, 85 to 235~s, and 202 to 330~s. The variations in the irradiation times were due to fluctuations in the X-ray flux between acquisitions, with the accumulated ionization chamber charges maintained constant in each case. For the ODD scan, the accumulated ionization chamber charges were adjusted between 0.28~nC and 1.13~nC to maintain approximately constant exposure at the image plate, accounting for inverse-square flux reduction and X-ray attenuation in air, corresponding to irradiation times between 212 and 272~s.

\begin{figure}[htbp]
    \centering
    \includegraphics[width=\linewidth]{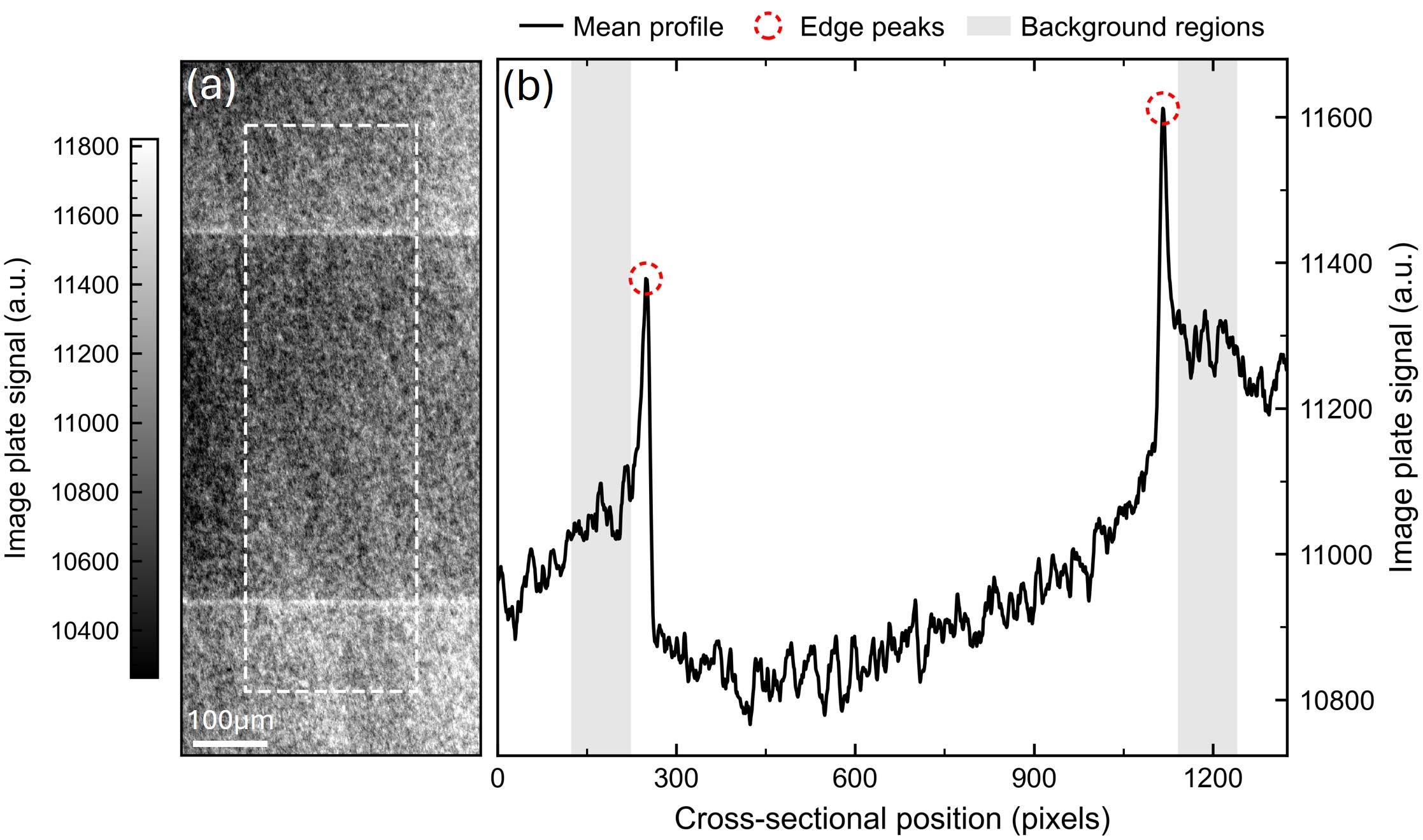}
    \caption{Example of the edge enhancement analysis for a single ROI. (a) Cropped image showing a portion of the nylon string, with edge enhancement visible at the nylon--air interfaces. The dashed rectangle indicates the selected ROI. (b) Mean cross-sectional profile showing the edge peaks and background regions used in the subsequent analysis.}
    \label{fig:edge_analysis}
\end{figure}

Phase-induced edge enhancement was quantified using cross-sectional intensity profiles across the nylon string. For each image, profiles were extracted from three regions of interest (ROIs) along the length of the string. Within each ROI, the two nylon--air interfaces were analysed independently, with each phase-induced edge peak compared with the mean intensity of 100 points in the corresponding background region outside the nylon string. The values obtained from the two interfaces were then averaged to obtain a single value for each ROI. The mean value across the three ROIs was reported for each imaging geometry. An example of the analysis procedure for a single ROI is shown in Fig.~\ref{fig:edge_analysis}.

For each interface, edge enhancement was calculated using Eq.~\ref{eq:edge_enhancement}, with $I_{\mathrm{bkg}}$ estimated as the mean intensity of 100 points in the corresponding background region. To account for the influence of image noise on the visibility of the edge enhancement signal, the peak intensity was additionally quantified relative to fluctuations in the corresponding local background using a z-score~\cite{illowsky2023introductory},

\begin{equation}
    z=\frac{I_E-I_{\mathrm{bkg}}}{\sigma_{\mathrm{bkg}}},
    \label{eq:zscore}
\end{equation}

where $\sigma_{\mathrm{bkg}}$ is the standard deviation of the intensity within the corresponding 100 point background region. The z-score provides a complementary measure of the visibility of phase-induced edge enhancement relative to background fluctuations. The same averaging procedure was applied to both metrics.

\subsection{Biological sample preparation and imaging}

Two wild-type zebrafish (\textit{Danio rerio}) at 20 days post fertilization (dpf) were used as biological imaging samples. The zebrafish were kept and raised under standard conditions~\cite{Westerfield2000} at the zebrafish facilities of the University of Santiago de Compostela (facility code: ES270280346401). The specimens were euthanized using an MS-222 (Sigma) overdose and fixed by immersion in 4\% paraformaldehyde in phosphate-buffered saline (PBS) at 4~$^\circ$C. Following washing in PBS, each specimen was mounted on a 125~$\mu$m thick polyethylene terephthalate (PET) film to provide mechanical support while minimizing the amount of surrounding material in the path of the X-ray beam. The two zebrafish were imaged at an SDD of 63~cm using geometric magnifications of $21\times$ and $42\times$, respectively. An irradiation time of 600~s was used for both acquisitions.

\section{Results and discussion}

\subsection{Experimental optimization of the PBI geometry}

\begin{figure}[htbp]
    \centering
    \includegraphics[width=\linewidth]{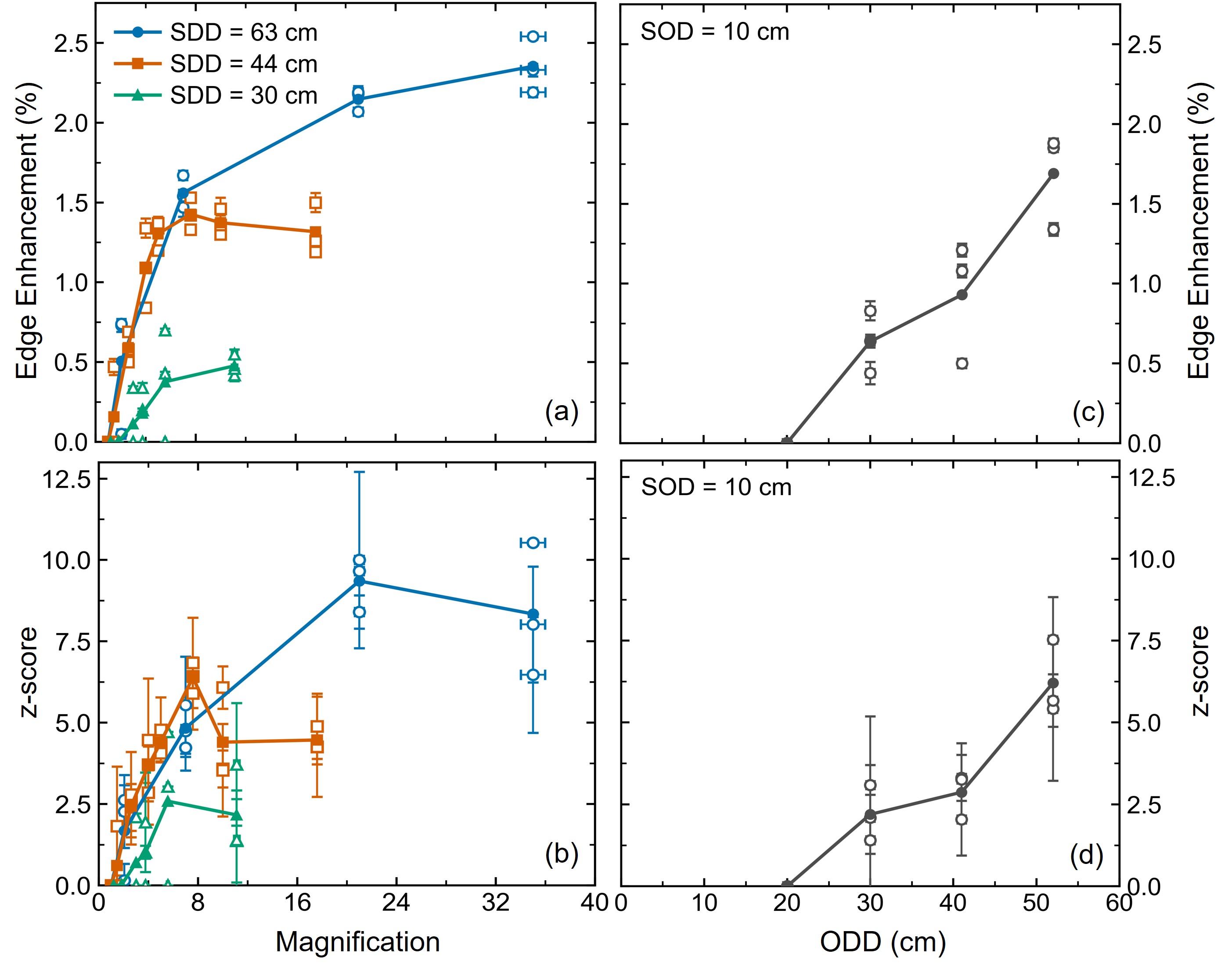}
    \caption{Experimental optimization of the PBI geometry using the nylon string phantom. (a,b) Edge enhancement and z-score as a function of geometric magnification for SDDs of 30, 44, and 63~cm. (c,d) Edge enhancement and z-score as a function of ODD at a fixed SOD of 10~cm. Open markers show the average values obtained from individual ROIs, each averaged across the two nylon–air interfaces. Filled markers connected by lines indicate the corresponding mean values across the three ROIs. Vertical error bars represent the standard error of the mean for each ROI. Horizontal error bars indicate positioning uncertainties.}
    \label{fig:geometry_optimization}
\end{figure}

The influence of the imaging geometry on the phase-induced edge enhancement was investigated experimentally using the nylon string phantom. Figure~\ref{fig:geometry_optimization}(a,b) shows the measured edge enhancement and supporting z-score metric as a function of geometric magnification for SDDs of 30, 44, and 63~cm. Figure~\ref{fig:geometry_optimization}(c,d) shows the measured edge enhancement and supporting z-score as a function of ODD for a fixed SOD of 10~cm.

As shown in Fig.~\ref{fig:geometry_optimization}(a), the edge enhancement generally increased with increasing SDD and magnification. At an SDD of 30~cm, weak to negligible edge enhancement was measured across the investigated magnification range. At an SDD of 44~cm, the edge enhancement increased rapidly with magnification before reaching a maximum at approximately $M=6\times$ and subsequently levelling off. At the largest SDD of 63~cm, the edge enhancement continued to increase over a broader range of magnifications before levelling off at approximately 2\% for magnifications above approximately $20\times$. The supporting z-score metric [Fig.~\ref{fig:geometry_optimization}(b)] showed a similar overall dependence on the imaging geometry, although with greater variability between the three ROIs.

The maximum accessible magnification was limited by the minimum achievable SOD of approximately 1.5~cm, required to prevent the nylon string phantom from obstructing the incident laser pulses or from interfering with the target positioning system. Importantly, such short SODs are achievable because the laser-driven X-ray source operates in ambient conditions, which reduces physical setup constraints around the X-ray generation region. Moreover, little reduction in edge enhancement was observed even at high geometric magnifications, suggesting that the micrometre-scale X-ray source size may help limit the effects of increasing source blur under these conditions.

The influence of ODD was additionally investigated at a fixed SOD of 10~cm. As shown in Fig.~\ref{fig:geometry_optimization}(c), the edge enhancement increased strongly with ODD, reaching a mean value of approximately 1.7\% at the maximum investigated ODD of 52~cm. Under these conditions, an ODD greater than approximately 20~cm was required to obtain measurable edge enhancement. The z-score [Fig.~\ref{fig:geometry_optimization}(d)] showed a similar trend, indicating that the edge enhancement signal became increasingly distinguishable from background fluctuations at larger ODDs.

The effect of imaging geometry on phase-induced edge enhancement has previously been investigated using laboratory X-ray sources. Dierks and Wallentin~\cite{dierks2020experimental} explored magnifications of up to approximately $1.6\times$ and $3.5\times$ using copper and tungsten microfocus X-ray tubes, respectively, in compact setups with corresponding SDDs of 12.8--48.7~cm and 5--10~cm. They identified optimal magnifications for edge enhancement of approximately $1.1\times$ and $1.47\times$, respectively. Dierks \textit{et al.}~\cite{Dierks:23} then subsequently investigated higher magnifications of $17$--$175\times$ using a nano-focus X-ray tube, with SDDs ranging from 14.3 to 34.3~cm, observing maximum edge enhancement at approximately $42\times$. The results presented here extend this approach to an imaging setup featuring a laser-driven X-ray source operating in ambient conditions, with magnifications between 1$\times$ and 35$\times$, and SDDs between 30~cm and 63~cm. While direct quantitative comparisons are difficult due to key differences in the X-ray source and the complicated detector response of the image plates, the results show a qualitatively similar trend. An initial increase in edge enhancement with magnification is seen, followed by a plateau at higher magnifications, rather than a gradual fall. This may suggest that the X-ray source can support even higher magnifications beyond those included in this study, although additional infrastructure, namely a larger optical table and precision sample manipulation stages, will be needed to investigate these higher magnifications.

The effect of imaging geometry on phase-induced edge enhancement has also been investigated using a laser-driven X-ray source. Gradl \textit{et al.}~\cite{gradl2017propagation} investigated the influence of ODDs up to approximately 1.6~m using an inverse-Compton X-ray source, observing increasing edge enhancement with ODD. These observations are qualitatively consistent with the present results, where edge enhancement also increased with ODD, although the maximum distance investigated was limited by the dimensions of the optical table.

\subsection{Effect of surrounding background material}

\begin{figure}[htbp]
    \centering
    \includegraphics[width=0.92\linewidth]{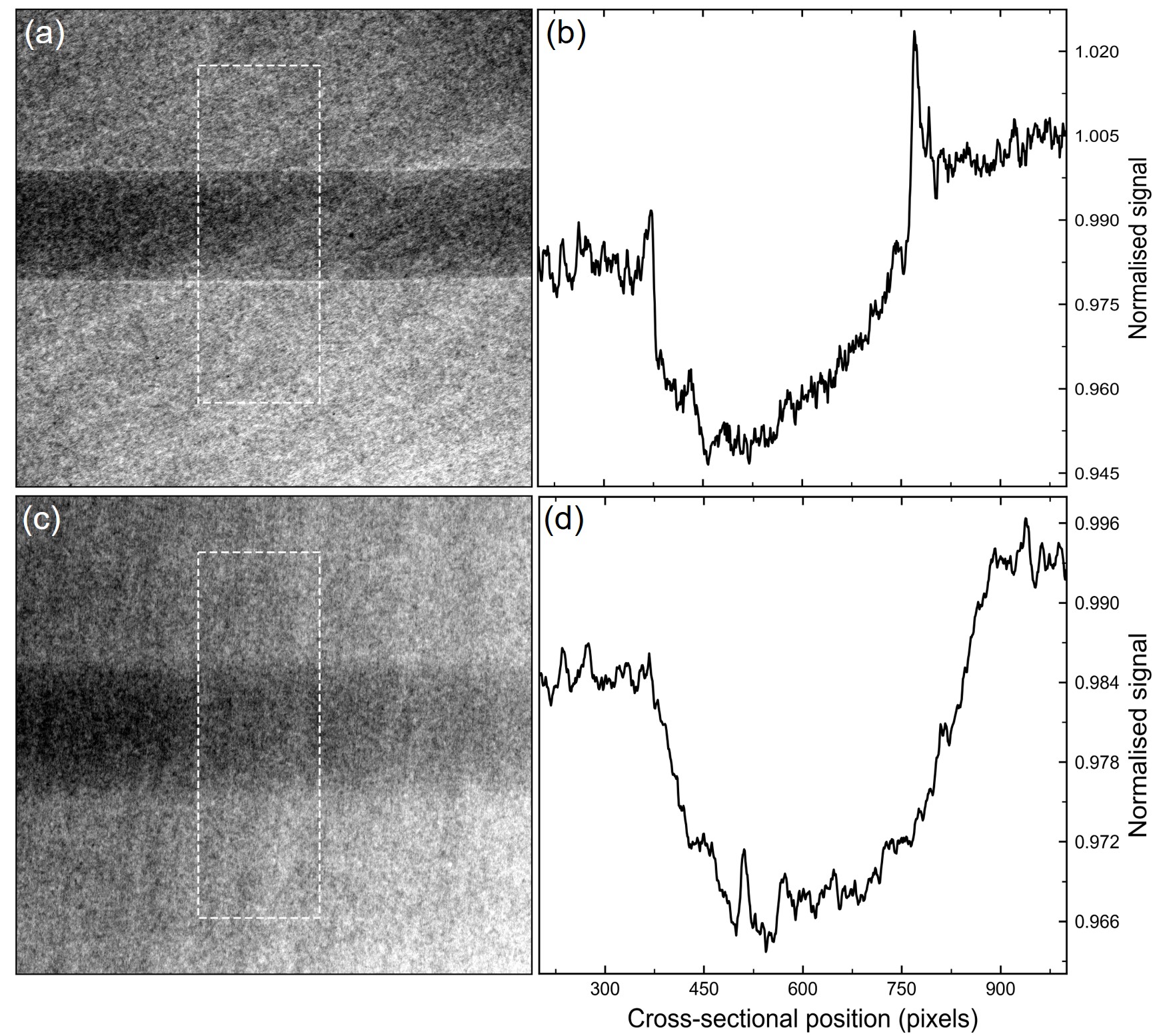}
    \caption{Effect of surrounding agar background material on the observed phase-induced edge enhancement. (a) Nylon string imaged in air and (b) corresponding normalised mean cross-sectional profile, showing edge enhancement features at the nylon--air interfaces. (c) Nylon string imaged with surrounding agar and (d) corresponding normalised mean cross-sectional profile, in which the edge enhancement features are strongly suppressed. Dashed rectangles indicate the ROIs used to obtain the profiles.}
    \label{fig:agar}
\end{figure}

To investigate the influence of the surrounding background material on the PBI capabilities of the laser-driven X-ray source, the nylon string phantom was imaged with and without surrounding agar under identical geometric conditions. An SOD of 3.6~cm and an SDD of 44~cm were used, corresponding to a geometric magnification of approximately $12.2\times$. Figure~\ref{fig:agar} presents a comparison between the nylon string imaged in air [Fig.~\ref{fig:agar}(a)] and with surrounding agar [Fig.~\ref{fig:agar}(c)]. The corresponding mean cross-sectional profiles are shown in Fig.~\ref{fig:agar}(b,d). At the nylon--air interfaces, edge enhancement is clearly visible as sharp intensity variations at the boundaries, particularly at the right interface where a strong peak is observed. In contrast, when the nylon string was surrounded by agar, the broad attenuation profile remained visible, while the sharp edge enhancement features associated with phase-contrast effects were strongly suppressed.

The reduction in edge enhancement in the presence of agar may result from a combination of the change in the material interface and the attenuation of X-rays by the agar. Replacing the nylon--air interface with a nylon--agar interface reduces the difference in the real part of the refractive index across the boundary, thus reducing the phase shift. In addition, lower-energy X-ray photons (such as the characteristic Cu K-$\alpha$ emission at approximately 8~keV~\cite{reija2025optimisation}) produce stronger phase effects than higher-energy X-ray photons, but are also more strongly attenuated by the surrounding material. Although it is difficult to distinguish the relative contributions of these effects from the present measurements, the observed reduction in edge enhancement shows the importance of considering the immediate surrounding material for sample preparation. However, it should be noted that in more complex samples, the strength of phase contrast will depend on the differences in refractive index at the various interfaces within the sample, which may include tissue--air boundaries as well as boundaries between neighbouring tissues.

\subsection{Application to biological imaging}

Following the experimental optimization of the PBI setup geometry, the capabilities of the laser-driven X-ray source for biological imaging were explored by imaging two zebrafish at 20~dpf. Based on the nylon string measurements, which showed increasing edge enhancement with SDD and sustained edge enhancement at high magnifications, the maximum accessible SDD of 63~cm was selected together with geometric magnifications of $21\times$ and $42\times$. The specific magnifications were chosen to provide suitable fields of view for resolving anatomical regions of interest, while also considering the physical constraints on sample positioning imposed by the experimental setup, particularly the microscope objective lens and target rotation system.

\begin{figure}[htbp]
    \centering
    \includegraphics[width=\linewidth]{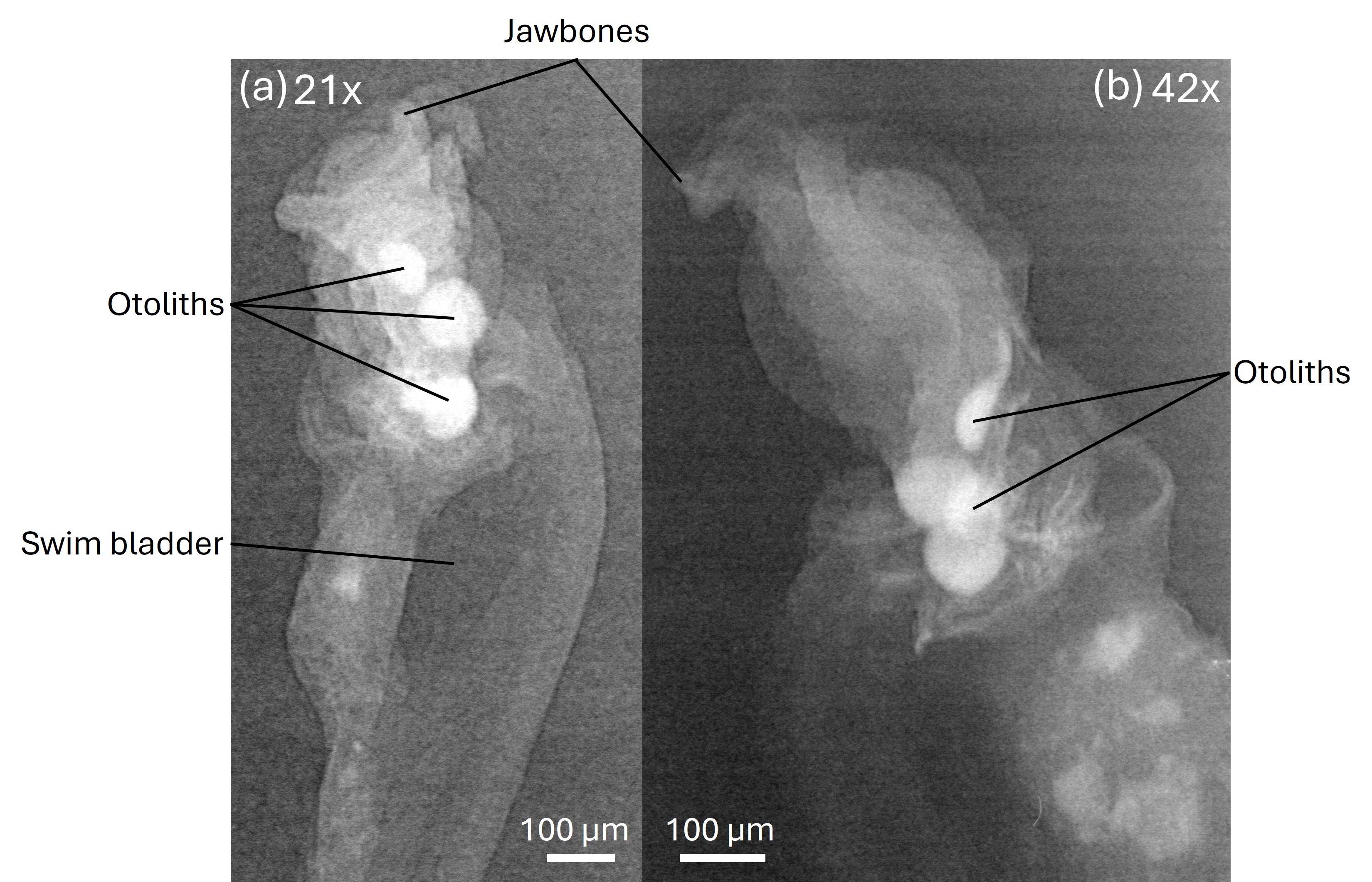}
    \caption{High-magnification propagation-based X-ray phase-contrast images of two 20~dpf zebrafish acquired at (a) $M=21\times$ and (b) $M=42\times$ geometric magnification. Phase-induced edge enhancement is visible along the external boundaries of the zebrafish and at multiple internal interfaces. Rostral is to the top in both images. Scale bars correspond to 100~$\mu$m.}
    \label{fig:zebrafish}
\end{figure}

Figure~\ref{fig:zebrafish} shows the resulting PBI images. At $M=21\times$ [Fig.~\ref{fig:zebrafish}(a)], the outline of the zebrafish and several internal structures, including the otoliths, jawbones, and swim bladder, are visible. Edge enhancement features consistent with propagation-based phase-contrast are visible along the external boundary of the sample and at several internal interfaces, such as around the jawbones and swim bladder. At $M=42\times$ [Fig.~\ref{fig:zebrafish}(b)], the increased magnification provides a more detailed view of the zebrafish head, with phase-induced edge enhancement remaining visible at multiple boundaries despite the large geometric magnification. The presence of edge enhancement features at these high magnifications is consistent with the results of the experimental optimization using the nylon string phantom. Moreover, the images demonstrate how the combination of a relatively large SDD and high geometric magnification can be applied to PBI of small biological samples. In particular, the micrometre-scale X-ray source helps preserve fine spatial details at high magnification, while operation in ambient conditions allows the short SODs required to achieve these large magnifications within a compact tabletop setup.

Previous studies have also applied X-ray phase-contrast imaging to biological samples using laser-driven X-ray sources. However, these studies have all featured systems requiring significantly higher laser powers (60-200~TW)~\cite{kneip2011x,cole2018high,wenz2015quantitative}, or otherwise more involved systems, such as an inverse-Compton X-ray source~\cite{gradl2017propagation}. A more directly comparable approach was reported by Gambari \textit{et al.}~\cite{gambari2020exploring}, who also demonstrated PBI of a biological sample using a laser-driven X-ray source featuring a solid metal (molybdenum) target. However, in that case the laser interaction occurred in a vacuum chamber, increasing the experimental setup complexity, with the maximum geometric magnification presented being 3.8$\times$. In contrast, the present source operates in ambient conditions using a nominal 30 GW laser, allowing shorter source-to-object distances and high geometric magnifications of up to 42× within a compact tabletop setup.

\section{Conclusion}

The suitability of a compact tabletop laser-driven X-ray source operating in ambient conditions for PBI has been investigated. The source, commissioned at the University of Santiago de Compostela, combines micrometre-scale X-ray source dimensions and high X-ray flux with the experimental flexibility afforded by stable X-ray generation without the need for a vacuum chamber.

The influence of imaging geometry on phase-induced edge enhancement was experimentally investigated using a nylon string phantom, providing practical guidance for PBI experiments with the laser-driven X-ray source. Increasing the SDD and the ODD produced a significant increase in edge enhancement, which was maintained at high geometric magnifications. At the maximum investigated SDD of 63~cm, an edge enhancement of approximately 2.2\% was measured at the highest tested geometric magnification of $35\times$.

The optimised conditions were subsequently applied to PBI of 20~dpf zebrafish at magnifications of $21\times$ and $42\times$, with phase-induced edge enhancement visible at external and internal boundaries. The micrometre-scale X-ray source size helps preserve fine spatial details at high magnification, while ambient operation permits the short SODs ($\sim$~1.5~cm) required to achieve these magnifications within a compact tabletop setup.

While these results demonstrate the potential for biological imaging, the suppression of edge enhancement observed when the nylon string was surrounded by agar shows the influence of surrounding material on phase-contrast visibility and is a consideration for sample preparation. Future work will focus on exploring alternative target materials to tailor the X-ray spectrum and integrating a permanent X-ray detector to further improve imaging performance. Additionally, motorised detector and sample stages will be incorporated into the setup to enable more advanced imaging experiments.

These results demonstrate the feasibility of high-magnification PBI using a compact solid-target X-ray source driven by a nominal 30~GW laser operating at 1~kHz. The combination of comparatively low laser-power requirements, micrometre-scale X-ray source size and operation in ambient conditions offers a promising alternative for laboratory-based X-ray phase-contrast imaging.

\section*{Acknowledgments}

The authors acknowledge financial support from Xunta de Galicia (ED431F2023/21, CIGUS Network of Research Centres), Agencia Estatal de Investigación (PID2023-152894OA-I00, RYC2021-032654I, CEX2023-001318-M), the Spanish Ministry of Science, Innovation and Universities and Agencia Estatal de Investigación (MICIU/AEI, PID2024-157431NB-C21), the Severo Ochoa Programme for Centres of Excellence (CEX2023-001292-S), and the Conselleria de Educación, Cultura y Universidades de la Generalitat Valenciana (CIPROM/2026/32).

\bibliography{references}

\end{document}